\documentclass[fleqn,11pt,twoside]{article}

\usepackage{amsthm,amsthm,amssymb, color, xcolor,epsfig, graphics, subfigure}

\usepackage{amsmath, graphicx, latexsym, lscape }

\makeatletter
\newcommand{\copyrightnote}[2]{{\renewcommand{\thefootnote}{}
 \footnotetext{\small\it
\begin{flushleft}
 \copyright \ #1   #2  
\end{flushleft}}}}

\newcommand{\Name}[1]{\begin{flushleft}
                       \LARGE \bf #1
                       \end{flushleft}\vspace{-3mm}}

\newcommand{\Author}[1]{\begin{flushleft}
                       \it #1 \end{flushleft}}

\newcommand{\Address}[1]{\begin{flushleft}
                       \it #1 \end{flushleft}}

\newcommand{\Date}[1]{\begin{flushleft}
                      \small  \it #1 \end{flushleft}}

\newcommand{\evenhead}{Author \ name}
\newcommand{\oddhead}{Article \ name}

\renewcommand{\@evenhead}{
\hspace*{-3pt}\raisebox{-15pt}[\headheight][0pt]{\vbox{\hbox to \textwidth
{\thepage \hfil \evenhead}\vskip4pt \hrule}}}
\renewcommand{\@oddhead}{
\hspace*{-3pt}\raisebox{-15pt}[\headheight][0pt]{\vbox{\hbox to \textwidth
{\oddhead \hfil \thepage}\vskip4pt\hrule}}}
\renewcommand{\@evenfoot}{}
\renewcommand{\@oddfoot}{}

\long\def\@makecaption#1#2{%
  \vskip\abovecaptionskip
  \sbox\@tempboxa{\small \textbf{#1.}\ \ #2}%
  \ifdim \wd\@tempboxa >\hsize
    {\small \textbf{#1.}\ \ #2}\par
  \else
    \global \@minipagefalse
    \hb@xt@\hsize{\hfil\box\@tempboxa\hfil}%
  \fi
  \vskip\belowcaptionskip}

\newcommand{\JNMPnumberwithin}[3][\arabic]{%
  \@ifundefined{c@#2}{\@nocounterr{#2}}{%
    \@ifundefined{c@#3}{\@nocnterr{#3}}{%
      \@addtoreset{#2}{#3}%
      \@xp\xdef\csname the#2\endcsname{%
        \@xp\@nx\csname the#3\endcsname .\@nx#1{#2}}}}%
}

\newcommand{\resetfootnoterule} {
  \renewcommand\footnoterule{%
  \kern-3\p@
  \hrule\@width.4\columnwidth
  \kern2.6\p@}
}

\renewcommand{\footnoterule}{}

\makeatother

\theoremstyle{definition}

\begin{document}

\renewcommand{\evenhead}{ {\LARGE\textcolor{blue!10!black!40!green}{{\sf \ \ \ ]ocnmp[}}}\strut\hfill 
E N Antonov and A Yu Orlov
}
\renewcommand{\oddhead}{ {\LARGE\textcolor{blue!10!black!40!green}{{\sf ]ocnmp[}}}\ \ \ \ \  
Sigma model instantons and singular tau function}

\thispagestyle{empty}
\newcommand{\FistPageHead}[3]{
\begin{flushleft}
\raisebox{8mm}[0pt][0pt]
{\footnotesize \sf
\parbox{150mm}{{Open Communications in Nonlinear Mathematical Physics}\ \  \ {\LARGE\textcolor{blue!10!black!40!green}{]ocnmp[}}
\ \ Vol.4 (2024) pp
#2\hfill {\sc #3}}}\vspace{-13mm}
\end{flushleft}}

\FistPageHead{1}{\pageref{firstpage}--\pageref{lastpage}}{ \ \ Article}

\strut\hfill

\strut\hfill

\copyrightnote{The author(s). Distributed under a Creative Commons Attribution 4.0 International License}

\Name{Sigma model instantons and singular tau function}

\Author{E. N. Antonov$^{\,1}$ and A. Yu. Orlov$^{\,2}$}

\Address{$^{1}$ Petersburg Nuclear Physics Institute named by B.P.Konstantinov of NRC ``Kurchatov Institute'',  1 mkr. Orlova Roshcha, Gatchina, Leningradskaya Obl, 188300, Russia,
email: antonov@thd.pnpi.spb.ru \\[2mm]
$^{2}$ 
Institute of Transmition Information Problems RAS, Moscow 127051 Russia, Bolshoy Karetny per. 19, build. 1;
Institute of Oceanology, Nahimovskii Prospekt 36,
Moscow 117997, Russia, email: orlovs@ocean.ru;
Kurchatov Institute, Moscow}

\Date{Received January 25, 2024; Accepted February 2, 2024}

\setcounter{equation}{0}

\begin{abstract}
\noindent 

The generating series for the instanton contribution to Green functions of the $2D$ sigma
model was found in the works of Schwarz, Fateev and Frolov. We show that this series
can be written as a formal tau function of the two-component KP
hierarchy.  
The higher times of the two-component tau function allow to consider various multiparameter insertions
into the instanton partition function,  therefore
 the tau function can be treated as the generating function for various correlators.
The construction can be generalized to the multicomponent case, which gives more parameters for the generating function of the correlators.
We call it formal singular tau function because this tau function is a sum where
each term is the infrared and ultraviolet divergent one exactly as the series found
by the mentioned authors. However, one can regularize each divergent term of this singular tau function in such a way that it is still a tau function.
Thus, we enlarge the families of tau functions to work with.

\end{abstract}

\,
\qquad\qquad\qquad\qquad\qquad In memory of Vladimir E. Zakharov
\,

\label{firstpage}


\section{Introduction}

The main purpose of this paper is to interpretate the contribution of
instantons in the Euclidean Green function of the O(3) non-linear $\sigma $
model (or the continuum classical Heisenberg ferromagnetic in two space
dimensions) in terms of tau functions of integrable hierarchies. This model
can be described by the action 
\begin{equation}
S=\frac{1}{2f}\int \sum\limits_{a=1}^{3}\left( \partial _{\mu }\sigma
^{a}\left( x\right) \right) ^{2}  \label{1}
\end{equation}%
where $\sigma^{a},\, a=1,2,3$ are the components of the unit vector:  $%
\sum\limits_{a=1}^{3}\sigma ^{a}\left( x\right) \sigma ^{a}\left( x\right) =1
$ ; $\mu =0,1.$

The model is similar to a Yang-Mills theory and possesses exact multi-instanton
solutions. The Euclidean Green functions can be represented in the form 
\begin{equation}
\frac{\int \phi \left( \sigma \right) \exp \left( -S\right)
\prod\limits_{x}d\sigma \left( x\right) }{\int \exp \left( -S\right)
\prod\limits_{x}d\sigma \left( x\right) }  \label{2}
\end{equation}%
Here $\phi \left( \sigma \right) $ is an arbitrary functional of $\sigma $.
If we parametrize $\sigma \left( x\right) $ with use of the complex function 
\begin{equation}  \label{stereo}
\omega(z) =\frac{\sigma ^{1}(z)+i\sigma ^{2}(z)}{1+\sigma ^{3}(z)}
\end{equation}
(the stereographic projection) obtained from the fields\ $\left( \sigma
^{1},\sigma ^{2},\sigma ^{3}\right) $\ and the complex variable $%
z=x_{0}+ix^{1}$ instead of the time and space coordinates $x_{0}$, $%
x_{1}$, then the instanton is the solution of the equation $\delta S=0$ with
the topological charge $q>0$ is given \cite{BP} 
\begin{equation}
\omega_q \left(a,b, z\right) =c\frac{\left( z-a_{1}\right) ...\left(
z-a_{q}\right) }{\left( z-b_{1}\right) ...\left( z-b_{q}\right) }  \label{3}
\end{equation}%
where $c$, $a_{i}$ and $b_{i}$ are arbitrary complex parameters.

Let us note that the classical $\sigma$-model in Minkowski space is the well-studied integrable model, see \cite{ZakharovNovikov-ed}.

\section{The instanton contribution and the $\protect\tau $ function}

In \cite{FFS}, the instanton contribution to the Euclidean Green
functions of the fields $\sigma$ using the steepest descent approximation was
obtained. If $\phi$ is a functional of the instanton fields $\omega$, then
the evaluation of the functional integral around the instanton vacuums
yields \cite{FFS} the answers written in form of multiple integrals over
instanton parameters:
\begin{equation}
\left< \phi \right>_{\mathrm{inst}}=\left[\frac{\sum_{q\geq 0}\frac{K^{q}}{%
(q!)^{2}}\int \phi(\omega_q ) \prod_{i<j\leq q}\frac{
|a_{i}-a_{j}|^{2}|b_{i}-b_{j}|^{2}}{ |a_{i}-b_{j}|^{2}|b_{i}-a_{j}|^{2}}%
\prod_{i=1}^{q} \frac{d^2 a_i d^2 b_i}{|a_i-b_i|^2}} {\sum_{q\geq 0}\frac{%
K^{q}}{(q!)^{2}}\int \prod_{i<j\leq q}\frac{
|a_{i}-a_{j}|^{2}|b_{i}-b_{j}|^{2}}{ |a_{i}-b_{j}|^{2}|b_{i}-a_{j}|^{2}}%
\prod_{i=1}^{q} \frac{d^2 a_i d^2 b_i}{|a_i-b_i|^2}} \right]_{\mathrm{reg}},
\label{4}
\end{equation}%
where $K$ is a real constant obtained as the result of the regularization
procedure\footnote{%
According to \cite{FFS} the constant $K$ is proportional to $k_0 f^{-2}_{%
\mathrm{phys}} \exp \left(-4\pi f^{-1}_{\mathrm{phys}}\right)\nu$ where $\nu$
is the substraction point, $f_{\mathrm{phys}}$ is a physical coupling
constant, $k_0$ is a constant depending on the cutoff method.}, and where for
each $q$ the instanton solution $\omega $ is given by (\ref{3}). The
denominator in (\ref{4}) coincides with the partition function $\Xi $\ of
the neutral classical two-dimensional Coulomb system (CCS) in the grand
canonical ensemble with the definite temperature T (T=1 see \cite{FFS})
(such a system was called the system of instanton quarks in \cite{FFS}). The
point T=1 is above the critical temperature (which is about T=1/2); this
means that the Coulomb gas is in the plasma phase. (Below the critical
temperature the Coulomb particles form dipoles). The symbol $[]_{\mathrm{reg}%
}$ means that this expression should be regularized in the ultraviolet limit,
where $a_i \to b_j$. Physical answers do not depend on the method of the
regularization.

Note that, in fact, instanton-anti-instanton interaction is also significant (see {%
Lipatov-Bukhvostov} \cite{LipatovBukhvostov}) but this was not considered in
the work  \cite{FFS}, and we also will not touch on this much more involved
topic.

\paragraph{Regularization.}

Let us notice that the answer (\ref{4}) was obtained \cite{FFS} as the
result of the calculation of the functional integral and a certain
regularization procedure and, in turn, the multiple integrals in (\ref{4})
are both infrared (IR) and ultraviolet (UV) divergent, and one needs an
additional regularization procedure. In short, it is discussed in \cite{FFS},
page 11.

As for the IR divergence (the divergence in the limit $a_i,b_i\to \infty$), 
it just means that one should be interested in the densities of the
instanton partition function (and of the correlation function) rather than
the partition and the correlation functions themselves. Then it is
reasonable to restrict the domain of the integration over each $a_i$ to the $%
D=L\times L$ box in the complex plane, the same for $b_i$ \cite{FFS}. To get
the density we divide each integral over $L^2$, simultaneously we send the
constant $K$ to $KL^2$. 

As for the ultraviolet regularization in the regions $b_i \approx a_j$ there
are different ways:

(A) We can do the following: we produce the replacement $b_i \to b_i
+\epsilon,\, {\bar b}_i \to {\bar b}_i -\epsilon$, where $\epsilon$ is a
small real number where $\epsilon^{-1}$ may be treated as a cutoff in the
momentum space.

In particular, for the one-instanton partition function, we get 
\begin{equation}  \label{UVreg-A}
K\int \frac{d^2a d^2 b}{|b-a|^2}\, \to \, \left(KL^2\right) L^{-2}\int_{D^2} 
\frac{d^2a d^2 b}{|a-b|^2_\epsilon }\,\quad\mathrm{where}\quad
|a-b|^2_\epsilon : =|a-b|^2 -\epsilon^2+i\epsilon \Im (a-b)
\end{equation}
The contribution of the region $b \approx a$ is finite and of order $%
\epsilon^{-1}$. Let us notice that, thanks to the structure of the numerators
inside the integrals in (\ref{4}), the order of the $q$-instanton integral
is $\epsilon^{-q}$. Thus, to get finite expressions, we send $K\to KL^2\epsilon$.

(B) One is to replace integrals by sums, that is, to consider the Coulomb gas
on the 2D lattice as mentioned in \cite{FFS} with the list of references. We
can do it as follows: we take a small real number $h$ (square grid spacing)
and set 
\begin{equation}  \label{a_nmb_nm}
a(n,m)=nh+imh\,\quad b(n,m)=(n+\gamma )h+i(m+\gamma^{\prime })h
\end{equation}
with non-integer $\gamma, \gamma^{\prime }$. In fact, we have two lattices:
one for positive and the other for negative Coulomb particles:
\begin{equation}  \label{UVreg-B}
K\int \frac{d^2a d^2 b}{|b-a|^2}\quad \to \quad \left(KL^2\right)
L^{-2}\sum_{0\le n,n^{\prime },m,m^{\prime }\le L} \frac{h^{-2}}{|n^{\prime
}-n+im^{\prime }-im+\frac 12(\gamma +i\gamma^{\prime 2 }}
\end{equation}
The summation range $0\le n,m \le L$ will also be denoted $D$, as in the
previous case.

Our goal is to relate (\ref{4}) with the regularizations (A)-(B) to
classical integrable systems.

\section{Tau functions}

\subsection{Two-sided two-component KP and the regularization (A)\label{2KP-regA}}

In this case, we use (\ref{UVreg-A}) and write the correlation function as 
\begin{equation}
\left< \phi \right>_{\mathrm{inst}}^A= \frac{\sum_{q\geq 0}\frac{K^{q}}{%
(q!)^{2}}\int_{D^{2q}} \phi(\omega_q ) \prod_{i<j\leq q}\frac{
|a_{i}-a_{j}|^{2}|b_{i}-b_{j}|^{2}} {|a_{i}-b_{j}|^{2}_\epsilon
|b_{i}-a_{j}|^{2}_\epsilon} \prod_{i=1}^{q} \frac{d^2 a_i d^2 b_i}{%
|a_i-b_i|^2_\epsilon } } {\sum_{q\geq 0}\frac{K^{q}}{(q!)^{2}}\int_{D^{2q}}
\prod_{i<j\leq q}\frac{ |a_{i}-a_{j}|^{2}|b_{i}-b_{j}|^{2}} {%
|a_{i}-b_{j}|^{2}_\epsilon |b_{i}-a_{j}|^{2}_\epsilon} \prod_{i=1}^{q} \frac{%
d^2 a_i d^2 b_i}{|a_i-b_i|^2_\epsilon }}  \label{4A}
\end{equation}
We will see that it is
a certain $\tau $ function of the two-sided two-component KP. 
Multi-component KP tau functions were introduced in the works of the Kyoto School \cite{JM}
in terms of free fermion formalism, see also
 later works \cite{Takasaki-Schur} and \cite{vanLeur}. The construction of tau functions
 implies the use of free massless fermions:
\begin{equation}
\psi ^{(\alpha )}(z)=\sum_{i\in \mathbb{Z}}\psi _{i}^{(\alpha )}z^{i},\quad
\psi ^{\dagger (\alpha )}(z)=\sum_{i\in \mathbb{Z}}\psi _{i}^{\dagger
(\alpha )}z^{-1-i}  \label{5}
\end{equation}%
where $\alpha $ is a ``color'' of fermions\ ($\alpha =1,2)$ and anti-commutation expressions for 
the Fermi modes $\psi _{i}^{(\alpha )},\psi _{i}^{\dagger (\alpha )}$\ are%
\begin{equation}
\left[ \psi _{i}^{(\alpha )},\psi _{j}^{(\beta )}\right] _{+}=0\qquad \left[
\psi _{i}^{\dagger (\alpha )},\psi _{j}^{\dagger (\beta )}\right] =0\qquad %
\left[ \psi _{i}^{(\alpha )},\psi _{j}^{\dagger (\beta )}\right] _{+}=\delta
_{\alpha ,\beta }\delta _{i,j}  \label{6}
\end{equation}%
The fermionic states with occupied levels up to $n^{(1)},n^{(2)}$  satisfy
the conditions%
\begin{equation*}
\langle n^{(1)},n^{(2)} | m^{(1)},m^{(2)} \rangle =\delta _{n^{(1)},m^{(1)}
}\delta _{n^{(2)},m^{(2)} }
\end{equation*}%
\begin{equation}
\psi _{i}^{(\alpha )}|n^{(\alpha )},*\rangle =\langle n^{(\alpha )},*|\psi
_{i}^{\dagger (\alpha )}= \psi _{-1-i}^{\dagger (\alpha )}|n^{(\alpha
)},*\rangle = \langle n^{(\alpha )},*|\psi _{-1-i}^{(\alpha )}=0,\quad
i<n^{(\alpha )}  \label{7}
\end{equation}%
We denote the sets $n^{(\alpha)},t^{(\alpha)}_{\pm 1},t^{(\alpha)}_{\pm 2},\dots$ by
${\bf t}^{(\alpha)}$, $\alpha=1,2$ where $n^{(\alpha)}$ are integers and where $t^{(\alpha)}_{\pm i}$ are complex parameters. The sets $\{t^{(\alpha)}_{\pm i},i>0\}$ we denote ${\bf t}^{(\alpha)}_\pm$. 

The family of tau functions of the two-sided two-component KP, which is related
to (\ref{4A}), is given by 
\begin{equation*}
\tau (n,{\bf t}^{(1)},{\bf t}^{(2)})= 
\end{equation*}
\begin{equation}\label{TSTCtau}
\langle n^{(1)},n^{(2)}|\,\Gamma_1 \left(
{\bf t}^{(1)}_+\right)\Gamma_2\left({\bf t}^{(2)}_+\right) g_1 g_2 
\Gamma^\dag_1 \left(
{\bf t}^{(1)}_-\right)\Gamma^\dag_2\left({\bf t}^2_-\right)
|n^{(2)}-n,n^{(1)}+n\rangle .  
\end{equation}%
where
\begin{equation}\label{g1g2}
g_1=e^{K^{\frac{1}{2}}\int_{D^2} \psi
^{(1)}(a)\psi ^{\dag (2)}({\bar{a}})d^2 a },\quad g_2=e^{K^{\frac{1}{2}}\int_{D^2}
\psi ^{(2)}({\bar{b}}-\epsilon)\psi ^{\dag (1)}(b+\epsilon)d^2 b
}
\end{equation}
where the ``evolution operators''
\begin{equation}
\Gamma_\alpha \left( {\bf t}^{(\alpha)}_\mp\right) =e^{\sum_{i>0} t_{ i}^{(\alpha
)}J_{i}^{(\alpha )}},\quad 
\Gamma_\alpha^\dag \left( {\bf t}^{(\alpha)}_\mp\right) =e^{\sum_{i>0} t_{- i}^{(\alpha)}J_{- i}^{(\alpha )}}
\label{9}
\end{equation}%
are expressed in terms of the modes of the currents $J_{i}^{(\alpha )}$. 
The current is given by 
\begin{equation}
:\psi ^{(\alpha )}(z)\psi ^{\dag (\alpha )}(z):=\psi ^{(\alpha )}(z)\psi
^{\dag (\alpha )}(z)-\langle 0|\psi ^{(\alpha )}(z)\psi ^{\dag (\alpha
)}(z)|0\rangle =\sum_{i\in \mathbb{Z}}\,J_{i}^{(\alpha )}\,z^{i-1}
\label{10}
\end{equation}%
It's modes can be written as follows:
\begin{equation}
J_{m}^{(\alpha )}=\sum_{i\in \mathbb{Z}}:\psi _{i}^{(\alpha )}\psi
_{i+m}^{(\alpha )}:  \label{11}
\end{equation}%
We have the Heisenberg algebra:
\begin{equation}
\left[ J_{k}^{(\alpha )},J_{m}^{(\beta )}\right] =k\delta _{\alpha ,\beta
}\delta _{k+m,0}\quad .  \label{12}
\end{equation}%
The discrete variables $n,n^{(\alpha)}$ and complex parameters $%
t^{(\alpha)}_{\pm i}$, $\alpha=1,2$, $i=1,2,3,\dots$ are called the higher times of the 
two-sided two-component KP hierarchy. In what follows, we put $n=0$ and omit it from the notations.

{\bf Remark}.
We use the term “two-sided” in relation to tau functions if we are interested in the dependence of the tau function on
both sets: on ${\bf t}^{(1,2)}_+$ and
by ${\bf t}^{(1,2)}_-$. Otherwise, we call it just a ``two-component'' tau function ($\alpha=1,2$).

{\bf Remark}.
The insertion of $g_1g_2$ can be interpreted as the introduction of mass for fermions, which were initially massless. This approach is developed in the next work \cite{AO3}.

As a result of direct evaluation (using the relations in Appendix \ref%
{useful-relations}), we obtain the following $\tau $ function of the two-sided
two-component KP: 
\begin{equation}
\tau^A \left(\mathbf{t}^{1},\mathbf{t}^{2}|D,\epsilon\right)= \sum_{q\geq 0}%
\frac{K^{q}}{(q!)^{2}}\int_{D^{2q}} \Phi _q\left({\bf a},{\bf b},\mathbf{t}^{1},\mathbf{t%
}^{2}\right) \prod_{i<j\leq q}\frac{ |a_{i}-a_{j}|^{2}|b_{i}-b_{j}|^{2}} {%
|a_{i}-b_{j}|^{2}_\epsilon |b_{i}-a_{j}|^{2}_\epsilon} \prod_{i=1}^{q} \frac{%
d^2 a_i d^2 b_i}{|a_i-b_i|^2_\epsilon }  \label{13}
\end{equation}%
where the function 
\begin{equation}\label{Phi}
\Phi _q\left({\bf a},{\bf b},\mathbf{t}^{1},\mathbf{t}^{2}\right)=\prod_{i=1}^{q} \left( 
\frac{a_i}{b_i}\right) ^{n^{(1)}}\left( \frac{\bar{a}_i}{\bar{b}_i}\right)
^{-n^{(2)}}e^{\theta(a_i,{\bf t}^1)-\theta(\bar{a}_i,{\bf t}^2) +\theta(b_i,{\bf t}^2)-\theta(\bar{b}_i,{\bf t}^1) }
\end{equation}
\begin{equation}\label{V}
\theta(z,{\bf t}^1) =V(z,{\bf t}_+^1)+V(z^{-1},{\bf t}_-^1),\quad
V(z,t)=\sum_{m>0}t_{m}z^{m}\quad .  
\end{equation}%
We imply that the parameters ${\bf t}^{(1,2)}$ are chosen in such a way that the integrals in (\ref{13}) are convergent.

Moreover, if we modify the dependence of $\theta$ on higher times, according to
\begin{equation}\label{V-modified}
\theta(z,{\bf t}) =V(z,{\bf t}_+)+V(z^{-1},{\bf t}_-)+\sum_{\alpha=1}^P V((z-s_\alpha,{\bf p}^{\alpha}))  ,\quad 
\end{equation}%
we obtain the tau function of the $(P+4)$-component tau function, where the additional sets of higher times
are the sets ${\bf p}^\alpha=(p_1^{(\alpha)},p_2^{(\alpha)},p_3^{(\alpha)},\dots),\,\alpha=1,\dots,P$.
In what follows, we will not use this additional freedom. In this short work, we shall use only the sets ${\bf t}^1_+$ and ${\bf t}^2_-$ which will be denoted ${\bf t}^1$ and ${\bf t}^2$, respectively.

Because $\Phi _q\left({\bf a},{\bf b},0,0\right)=1$, the tau function evaluated at 
${\bf t}^{1}={\bf t}^2=0$ is equal to the instanton grand
partition function $\tau^A \left(0,0|D,\epsilon\right) = Z_{\mathrm{inst}} $
and, for 
\begin{equation}
\phi_q ({\bf a},{\bf b})=\Phi_q\left({\bf a},{\bf b},\mathbf{t}^{1},\mathbf{t}^{2}\right)  \label{16}
\end{equation}
we observe 
\begin{equation}
\left< \phi \right>_{\mathrm{inst}}^A= \frac{\tau^A \left(\mathbf{t}^{1},%
\mathbf{t}^{2}|D,\epsilon\right)}{\tau^A \left(0,0|D,\epsilon\right)},
\label{17}
\end{equation}%

There is another and shorter way to write down the tau function (\ref{TSTCtau}). Let us introduce 
\begin{equation}
\psi^{(i)}(z,{\bf t}^i)=e^{\theta(z,{\bf t}^i)}\psi^{(i)}(z),\quad 
\psi^{\dag(i)}(\bar{z},{\bf t}^i)=e^{-\theta(\bar{z},{\bf t}^i)}\psi^{\dag(i)}(\bar{z}),\quad i=1,2
\end{equation}

Then, the multi-component tau function can be written as
\begin{equation}
\tau({\bf t},n^{(1)},n^{(2)})=c({\bf t})\langle n^{(1)},n^{(2)}|g_1({\bf t})g_2({\bf t})|n^{(1)},n^{(2)}\rangle
\end{equation}
where $c({\bf t})=\exp\sum_{i=1,2}\sum_{m>0} m t^{(i)}_mt^{(i)}_{-m}$ and where $g_i,\,i=1,2$ are
given by (\ref{g1g2}) where the Fermi fields 
$$\psi^{(1)}(z),\,\,\psi^{\dag(1)}(\bar{z}),\,\,\psi^{(2)}(z),\,\,\psi^{\dag(2)}(\bar{z})$$ 
are replaced respectively by 
$$\psi^{(1)}(z,{\bf t}^1),\,\,\psi^{\dag(1)}(\bar{z},\,\,{\bf t}^1),\,\,
\psi^{(2)}(z,{\bf t}^2),\,\,\psi^{\dag(2)}(\bar{z},{\bf t}^i)$$.

In the rest of the paper, we put $n^{(1)}=n^{(2)}=0$.


\paragraph{Discrete KP equations.}

If we specify the parameters as follows: 
\begin{equation}  \label{HMiwa}
t^{(1)}_k[\mathtt{n},z] :=-\frac 1k \sum_{i=1}^N \mathtt{n}_i z_i^{-k},\quad
t^{(2)}_k[\mathtt{m},y]=-\frac 1k \sum_{i=1}^M \mathtt{m}_i y_i^{-k}
\end{equation}
and denote such sets as $\mathbf{t}^{1}[\mathtt{n},z]$ and $\mathbf{t}^{2}[%
\mathtt{m},y]$, we obtain 
\begin{equation}  \label{Phi-omega}
\Phi_q\left(a,b,\mathbf{t}^{1}[\mathtt{n},z],\mathbf{t}^{2}[\mathtt{m}%
,y]\right) = \prod_{i=1}^N \left(\omega_q(a,b,z_i)\right)^{\mathtt{n}_i}
\prod_{i=1}^M \left(\omega_q({\bar a},{\bar b},y_i)\right)^{-\mathtt{m}_i}
\end{equation}
where $\omega_q$ was defined by (\ref{3}). Let us notice that $%
\omega(a,b,z)\omega({\bar a},{\bar b},{\bar z})=|\omega(a,b,z)|^2$.

The tau function written in the variables defined by (\ref{HMiwa}) solves the
so-called discrete KP equation; see \cite{JM} (and \cite{Zabrodin-2012} for
the review). If $\sigma^{(i)}(x),\,i=1,2,3$ are instanton solutions of form (%
\ref{3})-(\ref{stereo}), then, for the correlation function 
\begin{equation*}
G_{\mathtt{n}_1,\mathtt{n}_2,\mathtt{n}_3}(z_1,z_2,z_3):= 
\end{equation*}
\begin{equation}
\left< \left(%
\frac{\sigma ^{1}(z_1)+i\sigma ^{2}(z_1)}{1+\sigma ^{3}(z_1)}\right)^{%
\mathtt{n}_1}  \left(\frac{\sigma ^{1}(z_2)+i\sigma ^{2}(z_2)}{1+\sigma
^{3}(z_2)}\right)^{\mathtt{n}_2}  \left(\frac{\sigma ^{1}(z_3)+i\sigma
^{2}(z_3)}{1+\sigma ^{3}(z_3)}\right)^{\mathtt{n}_3}  \right>_{\mathrm{inst}%
}^A 
\end{equation}
one can write the discrete Hirota bilinear equation (in other words, as the discrete
KP equation): 
\begin{equation*}
(z_2-z_3)G_{\mathtt{n}_1+1,\mathtt{n}_2,\mathtt{n}_3}(z_1,z_2,z_3)G_{\mathtt{%
n}_1,\mathtt{n}_2+1,\mathtt{n}_3+1}(z_1,z_2,z_3) 
\end{equation*}
\begin{equation*}
+(z_3-z_1)G_{\mathtt{n}_1,\mathtt{n}_2+1,\mathtt{n}_3}(z_1,z_2,z_3)G_{%
\mathtt{n}_1+1,\mathtt{n}_2,\mathtt{n}_3+1}(z_1,z_2,z_3) 
\end{equation*}
\begin{equation}  \label{Hirota-for-G}
+(z_1-z_2)G_{\mathtt{n}_1,\mathtt{n}_2,\mathtt{n}_3+1}(z_1,z_2,z_3)G_{%
\mathtt{n}_1+1,\mathtt{n}_2+1,\mathtt{n}_3}(z_1,z_2,z_3) = 0
\end{equation}
Other sets of equations may be written for general correlation functions
involving (\ref{Phi-omega}) (this will be done in a more detailed text).


\paragraph{Densities.}

As we mentioned, the denominator in (\ref{4}) coincides with the partition
function $\Xi $\ of the neutral classical Coulomb system (CCS) in the grand
canonical ensemble with the definite temperature T (T=1 see \cite{FFS}). 
\begin{equation}
\tau (0,0,0,0)=\Xi
\end{equation}%
The constant K plays the role of fugacity in the Coulomb system. The
expression (\ref{17}) also coincides with the correlation function of the
CCS (at T=1). Let us consider the instanton contribution $G^{\mathrm{inst}%
}\left( x,y\right) $ in the Green function%
\begin{equation}
G\left( x,y\right) =\langle \bigtriangleup _{x}\log |\omega \left( x\right)
|,\bigtriangleup _{y}\log |\omega \left( y\right) |\rangle  \label{20}
\end{equation}%
corresponding to functional $\phi \left( \omega \right) =\bigtriangleup
_{x}\log |\omega \left( x\right) |\bigtriangleup _{y}\log |\omega \left(
y\right) |$ that is $\rho (x)\rho (y)$ with

$\rho (x)=2\pi \left( \sum_{i}\delta \left( x-a_{i}\right) -\sum_{i}\delta
\left( x-b_{i}\right) \right) $. In order to obtain this result in terms of $%
\tau $ functions, we have to make the Miwa transformation of times%
\begin{equation}
t_{n}^{(\alpha )}=-\frac{t}{nx^{n}}-\frac{t}{ny^{n}}  \label{21}
\end{equation}%
then we achieve 
\begin{equation}
\bigtriangleup _{x}\bigtriangleup _{y}\frac{\partial }{\partial t}\left(
\prod_{i}^{q}\Phi _{0,0}(a_{i},b_{i},{\bf t}^1,{\bf t}^2)|_{t_{n}^{(\alpha )}
= -%
\frac{t}{nx^{n}}-\frac{t}{ny^{n}}}\right) |_{t=0}
\end{equation}
\begin{equation}
=
\bigtriangleup _{x}\log
|\omega \left( x\right) |\bigtriangleup _{y}\log |\omega \left( y\right)
|=\rho (x)\rho (y).  \label{22}
\end{equation}%
One can interpret $\rho (x)$ as the charge density. We see 
\begin{equation}
G^{\mathrm{inst}}\left( x,y\right) =\langle \bigtriangleup _{x}\log |\omega
\left( x\right) |\bigtriangleup _{y}\log |\omega \left( y\right) |\rangle
_{inst}=\langle \rho (x)\rho (y)\rangle _{CCS}  \label{23}
\end{equation}%
\begin{equation}
=\frac{C\bigtriangleup _{x}\bigtriangleup _{y}\frac{\partial }{\partial t}%
\left( \tau (0,0,{\bf t}^1,{\bf t}^2)|_{t_{n}^{(\alpha )}=-\frac{t}{nx^{n}}-%
\frac{t}{ny^{n}}}\right) |_{t=0}}{\tau (0,0,0,0)}  \label{24}
\end{equation}%
Similarly to the previous way, we can obtain the instanton contribution in
the more general Green function corresponding to the functional 
\begin{equation}
\phi \left( \omega \right) =\bigtriangleup _{x_{1}}\log |\omega \left(
x_{1}\right) |\bigtriangleup _{x_{2}}\log |\omega \left( x_{2}\right)
|...\bigtriangleup _{x_{m}}\log |\omega \left( x_{m}\right) |  \label{25}
\end{equation}%
by 
\begin{equation}
G^{\mathrm{inst}}\left( x_{1},x_{2},...x_{m}\right) =\langle \rho
(x_{1})\rho (x_{2})...\rho (x_{m})\rangle _{CCS}=  \label{26}
\end{equation}%
\begin{equation*}
=\frac{C\bigtriangleup _{x_{1}}\bigtriangleup _{x_{2}...}\bigtriangleup
_{x_{m}}\frac{\partial }{\partial t}\left( \tau
(0,0,{\bf t}^1,{\bf t}^2)|_{t_{n}^{(\alpha )}=-\frac{t}{nx_{1}^{n}}-\frac{t}{%
nx_{2}^{n}}...\frac{t}{nx_{m}^{n}}}\right) |_{t=0}}{\tau (0,0,0,0)}
\end{equation*}

\subsection{Two-component KP and the regularization (B)\label{2KP-regB}}

In this case, we have 
\begin{equation*}  
\left<\phi \right>_{\mathrm{inst}}^B=
\end{equation*}
\begin{equation}\label{4B}
\frac{ \sum_{q\geq 0}\frac{K^{q}}{%
(q!)^{2}}\sum_{D^{2q}} \phi _q\left(a,b\right) \prod_{i<j\leq q}\frac{
|a_{n_im_i}-a_{n_jm_j}|^{2}|b_{n_im_i}-b_{n_jm_j}|^{2}} {%
|a_{n_im_i}-b_{n_jm_j}|^{2} |b_{n_im_i}-a_{n_jm_j}|^{2}} \prod_{i=1}^{q} 
\frac{1}{|a_{n_im_i}-b_{n_im_i}|^2 }} {\sum_{q\geq 0}\frac{K^{q}}{(q!)^{2}}%
\sum_{D^{2q}} \prod_{i<j\leq q}\frac{
|a_{n_im_i}-a_{n_jm_j}|^{2}|b_{n_im_i}-b_{n_jm_j}|^{2}} {%
|a_{n_im_i}-b_{n_jm_j}|^{2} |b_{n_im_i}-a_{n_jm_j}|^{2}} \prod_{i=1}^{q} 
\frac{1}{|a_{n_im_i}-b_{n_im_i}|^2 }}
\end{equation}
where $\sum_{D^{2q}}$ means $\sum_{n_1,\dots, n_q,m_1,\dots m_q \in D}$ and
where $a_{nm}, b_{nm}$ are given by (\ref{a_nmb_nm}).

One just needs to replace integrals by sums according to (\ref{UVreg-B}) in
the expression (\ref{TSTCtau}): 
\begin{equation}
\tau_{2KP} (n^{(0)},n^{(1)},n^{(2)},{\bf t}^1,{\bf t}^2|D,h)= 
\langle n^{(1)},n^{(2)}|\,\Gamma \left(
{\bf t}^1\right)\Gamma\left({\bf t}^2\right) \,g\,|n^{(2)}-n^{(0)},n^{(1)}+n^{(0)}\rangle  \label{8A}
\end{equation}%
where
\begin{equation}
 g=e^{K^{\frac{1}{2}}\sum_{(k,m)\in
D^2} \psi ^{(1)}(a_{km})\psi ^{\dag (2)}({\bar a_{km}}) }\,e^{K^{\frac{1}{2}%
}\sum_{(k,m)\in D^2} \psi ^{(2)}({\bar b_{km}})\psi ^{\dag (1)}(b_{km})}
\end{equation}
and
where the summation range in the exponents is chosen as $0\le n,m \le L$. We
obtain the enumerator in (\ref{4B}): 
\begin{equation}
\tau^B \left(\mathbf{t}^{1},\mathbf{t}^{2}|D,h\right)= 
\end{equation}
\begin{equation}
\sum_{q\geq 0}\frac{%
K^{q}}{(q!)^{2}}\sum_{D^{2q}} \Phi _q\left(a,b,\mathbf{t}^{1},\mathbf{t}%
^{2}\right) \prod_{i<j\leq q}\frac{
|a_{n_im_i}-a_{n_jm_j}|^{2}|b_{n_im_i}-b_{n_jm_j}|^{2}} {%
|a_{n_im_i}-b_{n_jm_j}|^{2} |b_{n_im_i}-a_{n_jm_j}|^{2}} \prod_{i=1}^{q} 
\frac{1}{|a_{n_im_i}-b_{n_im_i}|^2 }  \label{13B}
\end{equation}%
If we choose $\phi_q(a,b)=\Phi_q(a,b,\mathbf{t}^1,\mathbf{t}^2)$ we get the
same relations as in the previous case A, we replace $\left< *\right>^A_{%
\mathrm{inst}}$ by $\left< *\right>^B_{\mathrm{inst}}$.

\subsection{One-component KP and the regularization (B)\label{KP-regB}}

The regularization (B) can also be written as the following KP tau function:
\begin{equation}  \label{8A'}
\tau^{\mathrm{B}}_{\mathrm{KP}}(\mathbf{t}|D,h)= \langle
n|\Gamma(t)e^{K\sum_{D^2}\frac{\psi(a_{nm})\psi^\dag(b_{nm})}{{\bar a}_{nm}-{%
\bar b}_{nm}}}|n\rangle
\end{equation}
where $a_{nm}$ and $b_{nm}$ are given by (\ref{a_nmb_nm}), and 
\begin{equation*}
\Gamma(t)=e^{\sum_{m>0} t_m J_m},\quad J_m=\sum_{i\in\mathbb{Z}}\psi_i
\psi^\dag_{i+m} 
\end{equation*}
$\Gamma(t)$, the Fermi fields and $\Phi_q$ are the same as in subsection \ref%
{2KP-regA}, where the second component is absent: 
\begin{equation*}
\Phi _q\left(a,b,\mathbf{t}\right)=\prod_{i=1}^{q} \left( \frac{a_i}{b_i}%
\right) ^{n}e^{V(a_i,t)-V(b_i,t) } ,\quad \Phi _q\left(a,b,\mathbf{t}[%
\mathtt{n},z]\right)=\prod_i \left( \omega(z_i) \right)^{\mathtt{n}_i} \label%
{Phi-KP} 
\end{equation*}
where $V$ was defined in (\ref{V}) and where $\mathbf{t}[\mathtt{n},z]$ denotes
the choice $n=0$ and $t_m=-\sum \mathtt{n}_i z_i^m,\,m>0$. We get the same
equation (\ref{Hirota-for-G}) for 
\begin{equation*}
G_{\mathtt{n}_1,\mathtt{n}_2,\mathtt{n}_3}(z_1,z_2,z_3):=
\left<\left(\omega(z_1)\right)^{\mathtt{n}_1}\left(\omega(z_2)\right)^{%
\mathtt{n}_2} \left(\omega(z_3)\right)^{\mathtt{n}_3} \right>_{\mathrm{inst}%
}^B 
\end{equation*}
Formula (\ref{8A'}) yields the same answer as (\ref{8A}) if we put $\mathbf{t%
}^2=0$ and $\mathbf{t}^1=\mathbf{t}$ (see Appendix \ref{useful-relations}).
However, in the case of the one-component KP, we can not construct $|\omega(z)|
$ by specializing the parameters $t_1,t_2,\dots$ in $\Phi$.

\subsection{Regularization C}

Few words about different regularization (without details).
This regularization is performed directly in the expression for the tau function.

(a) The ultri-violet regularization is achieved by the cutting out of higher Fermi modes:
\begin{equation}
 \psi^{(i)}(z)\,\to \,\psi^{(i)}(z;M)=\sum_{j\le M} z^j\psi_j^{(i)},\quad 
\psi^{\dag(i)}(z)\,\to \,\psi^{\dag(i)}(z;M)=\sum_{j\le M} z^j\psi_{-j-1}^{\dag(i)},
\end{equation}
where $M$ is the cutting parameter.

(b) The infra-red regularization is done via the including of the decay to the measure

\begin{equation}
 \int \psi^{(1)}(a)\psi^{(2)\dag}(\bar{a})d^2 a\,\to\,
\int \psi^{(1)}(a)\psi^{(2)\dag}(\bar{a})e^{-\epsilon |a|^2}d^2 a
\end{equation}

\section{Appendix. Useful relations\label{useful-relations}}

We use the following relations:
\[
 \Gamma(t)\psi(z)=e^{V(z,t)}\psi(z)\Gamma(t),\quad
 \Gamma(t)\psi^\dag(z)=e^{-V(z,t)}\psi^\dag(z)\Gamma(t)
\]
and $\Gamma(t)|n\rangle =|n\rangle$. Then
\[
 \langle n|\psi(z_1)\psi^\dag(y_1)\cdots \psi(z_q)\psi^\dag(y_q)|n\rangle =
\prod_{i<j}^q \frac{(z_i-z_j)(y_i-y_j)}{(z_i-y_j)(y_i-z_j)}\prod_{i=1}^q
\frac{1}{z_i-y_i}\left(\frac{z_i}{y_i}\right)^n
\]
Also
\[
 e^{\sum_{i,j}\xi_i\eta_jA_{i,j}}=
 1+\sum_{q>0}\sum_{\alpha_1>\cdots\alpha_q\atop \beta_1>\cdots >\beta_q}
 \xi_{\alpha_1}\cdots \xi_{\alpha_q}\eta_{\beta_1}\cdots
 \eta_{\beta_q} \det\left(A_{\alpha_i,\beta_j} \right)
\]
where $\xi_i,\eta_i$ are odd variables (Fermi fields with the property 
$\xi_i\eta_j+\eta_j\xi_i
 =0$
for each pair $i,j$),
and $A_{i,j}$ is a (possibly infinite) matrix.

And at last
\[
 \det\left(\frac{1}{z_i-y_j}  \right)=
\prod_{i<j} \frac{(z_i-z_j)(y_i-y_j)}{(z_i-y_j)(y_i-z_j)}\prod_{i}
\frac{1}{z_i-y_i}
\]

\section{Appendix. Bilinear identity for the two-component $\ \protect\tau $ function}

In this section we define more general $\tau $ functions in comparison with (%
\ref{TSTCtau}): 
\begin{equation}
\tau (n_{1},n_{2},n,{\bf t}^1,{\bf t}^2)=\langle n_{1},n_{2}|\,\Gamma \left(
{\bf t}^1,{\bf t}^2\right) g\,|n_{2}-n,n_{1}+n\rangle \quad ,  \label{27}
\end{equation}%
where 
\begin{equation}
g=\,e^{\int \sum_{i,j=1,2}:\psi ^{(i)}(a^{(i)}_1)\psi ^{\dag (j)}(a_2^{(j)}
): d\mu(a_1^{(i)},a_2^{(j)}) }\quad  \label{28}
\end{equation}%
where $:\psi ^{(i)}(a^{(i)}_1)\psi ^{\dag (j)}(a_2^{(j)} ): =\psi
^{(i)}(a^{(i)}_1)\psi ^{\dag (j)}(a_2^{(j)} ) - \langle 0,0|\psi
^{(i)}(a^{(i)}_1)\psi ^{\dag (j)}(a_2^{(j)} )|0,0\rangle $ while $\Gamma
\left( {\bf t}^1,{\bf t}^2\right) $ is given by (\ref{9}). Particularly 
interesting for us $\tau $ the following one: 
\begin{equation*}
\tau (n_{1},n_{2},{\bf t}^1,{\bf t}^2):=\tau (n_{1},n_{2},0,{\bf t}^1,{\bf t}^2)
\end{equation*}

The bilinear identity is valid in the following form (see \cite{JM}). For $%
n_{1}-n_{1}^{\prime }\geq n^{\prime }-n\geq n_{2}^{\prime }-n_{2}+2$, we have%
\begin{equation}
\sum_{\alpha =1}^{2}\oint \frac{dz}{2\pi iz}\langle n_{1},n_{2}|\,\Gamma
\left( {\bf t}^1,{\bf t}^2\right) \psi ^{(\alpha
)}(z)g\,|n_{2}-n-1,n_{1}+n\rangle  \label{29}
\end{equation}%
\begin{equation*}
\times \langle n_{1}^{\prime },n_{2}^{\prime }|\,\Gamma \left( t^{^{\prime
}(1)},t^{\prime (2)}\right) \psi ^{\dag (\alpha )}(z)g\,|n_{2}^{\prime
}-n^{\prime }+1,n_{1}^{\prime }+n^{\prime }\rangle =0
\end{equation*}%
and the integration is taken along a small contour at $z=\infty $ so that $%
\oint \frac{dz}{2\pi iz}=1$.

Rewriting this ($\ref{21}$), we obtain 
\begin{equation}
\oint \frac{dz}{2\pi iz}\left( -1\right) ^{n_{2}+n_{2}^{\prime
}}z^{n_{1}-1-n_{1}^{\prime }}e^{V(z,{\bf t}^1-t^{\prime (1)})}  \label{30}
\end{equation}%
\begin{equation*}
\times \tau (n_{1}-1,n_{2},n+1,{\bf t}^1-\theta \left( z^{-1}\right)
,{\bf t}^2)\tau (n_{1}^{\prime }+1,n_{2}^{\prime },n^{\prime }-1,t^{\prime
(1)}+\theta \left( z^{-1}\right) ,t^{\prime (2)})
\end{equation*}%
\begin{equation*}
+\oint \frac{dz}{2\pi iz}z^{n_{2}-1-n_{2}^{\prime }}e^{V(z,{\bf t}^2-t^{\prime
(2)})}
\end{equation*}%
\begin{equation*}
\times \tau (n_{1},n_{2}-1,n,{\bf t}^1,{\bf t}^2-\theta \left( z^{-1}\right)
)\tau (n_{1}^{\prime },n_{2}^{\prime }+1,n^{\prime },t^{\prime
(1)},t^{\prime (2)}+\theta \left( z^{-1}\right) )\quad ,
\end{equation*}%
where $\theta \left( z^{-1}\right) =\left( \frac{1}{z},\frac{1}{2z^{2}},...%
\frac{1}{nz^{n}},...\right) $.

As an example of (\ref{22}),  for 
$$
f= \tau (n_{1},n_{2},0,{\bf t}^1,{\bf t}^2) = \tau (n_{1},n_{2},{\bf t}^1,{\bf t}^2),$$
$$
g=\tau (n_{1}-1,n_{2}+1,1,{\bf t}^1,{\bf t}^2)
$$  
and 
$$
g^{\ast }=\tau(n_{1}+1,n_{2}-1,-1,{\bf t}^1,{\bf t}^2)
$$
we get the following bilinear equations:
\begin{equation}
\left( D_{t_{2}^{\left( 1\right) }}-D_{t_{1}^{\left( 1\right) }}^{2}\right)
f\cdot g=0,\qquad \left( D_{t_{2}^{\left( 1\right) }}-D_{t_{1}^{\left(
1\right) }}^{2}\right) g^{\ast }\cdot f=0,  \label{31}
\end{equation}%
\begin{equation*}
\left( D_{t_{2}^{\left( 2\right) }}+D_{t_{1}^{\left( 2\right) }}^{2}\right)
f\cdot g=0,\qquad \left( D_{t_{2}^{\left( 2\right) }}+D_{t_{1}^{\left(
2\right) }}^{2}\right) g^{\ast }\cdot f=0,
\end{equation*}%
\begin{equation*}
D_{t_{1}^{\left( 1\right) }}D_{t_{1}^{\left( 2\right) }}f\cdot f-2g\cdot
g^{\ast }=0,\qquad \qquad \qquad \qquad \qquad \qquad
\end{equation*}%
where Hirota operator is $D_{x}\sigma \cdot \tau =\frac{\lim }{%
\varepsilon \rightarrow 0}\frac{\partial }{\partial \varepsilon }\sigma
\left( x+\varepsilon \right) \tau \left( x-\varepsilon \right) =\sigma
_{x}\tau -\sigma \tau _{x}$

\section{Discussion} 

We have shown that the Fateev-Frolov-Schwartz instanton series coincides with the formal series for the tau function of the two-component KP hierarchy. This is a formal expression, the manipulation of which nevertheless has a physical meaning. Such a tau function can be treated in the same way as with expressions of quantum field theory - eliminated by divergences by introducing trims, by replacing the measure, by switching to a lattice theory. All this can be done using the fermion representation for the tau function.
In this case, based on the formal expression, one can obtain various equations for the correlation functions.
  However, in order to calculate the correlation functions themselves it is necessary to use not massless, but massive fermions, which obey the massive Dirac equation. 
  Note that this is related to works about the quantum model of $\sin$-Grodon and the Thirring model \cite{Pogreb}.
  
  This is what we are doing in our next work.
  
  Note that new connections with the theory of classical integrable systems appear here. Firstly, these are classic works of the Kyoto School \cite{Sato}.
  Then it is related to the so-called $\bar{\partial}$-problem for the KP equation \cite{ZahMan},\cite{Dbar}.

\subsection*{Acknowledgements}

The work of A.O. has been supported by the Russian Science
Foundation (Grant No.20-12-00195).

\label{lastpage}

\begin{thebibliography}{99}


\bibitem{BP} A.A. Belavin and A.M. Polyakov,  Metastable
states of two-dimensional isotropic ferromagnets, {\it Pis'ma
Zh. Eksp. Teor. Fiz.}, 1979, v. 22, N10,  503--506

\bibitem{FFS} V.A. Fateev, I.V. Frolov, A.S. Schwarz, Quantum Fluctuations
of Instantons in the Nonlinear $\sigma$ model, {\it Nuclear Physics B}, 1979 V.
154, N1, 1--20

\bibitem{LipatovBukhvostov} A.P. Bukhvostov and L.N. Litpatov,
Instanton-antiinstanton interaction in nonlinear $\sigma$-model and
certain exactly solvable fermionic theory, {\it Pis'ma v ZhETF} 1980, V.31, N 2
 138--142 (in Russian)

\bibitem{JM} M. Jimbo, T. Miwa, Solutons and Infinite
Dimensional Lie Algebra, {\it Publ. RIMS Kyoto Univ.} 1983, V.19 943--1000

\bibitem{vanLeur}  V. G. Kac, J. W. van de Leur, ``The $n$-component KP hierarchy and representation theory'',
{\it J.Math.Phys.} 2003, V.44, 3245-3293 ;
V. Kac, J. van de Leur, 
Multicomponent KP type hierarchies and their reductions, associated to conjugacy classes of Weyl groups of classical Lie algebras, {\it Journal of Mathematics and Physics} 2023, V.64, 091702,
arXiv:2304.05737 

\bibitem{Takasaki-Schur} K, Takasaki, Initial value problem for the Toda lattice hierarchy, {\it Adv. Stud. Pure Math.}, 1984, V.4, 139--163

\bibitem{Zabrodin-2012} A. Zabrodin, Bethe ansatz and Hirota equation in
integrable models, arXiv:1211.4428

\bibitem{ZakharovNovikov-ed} S.V. Manakov, S. P. Novikov, L. Pitaevski and
V. E. Zakharov, Theory of solitons, Nauka, Moscow 1979

\bibitem{AO3} E.N. Antonov, A.Yu. Orlov, Fateev-Frolov-Schwarz instanton sum and 
regularized tau function, to be published

\bibitem{Pogreb}  A. K. Pogrebkov, V. N. Sushko, Quantization of the $(\sin\psi)_2$ interaction in terms of fermion variables, {\it Theoretical and Mathematical Physics}, 1975, V.24, N 3, 935--937

\bibitem{Sato} M. Sato, T. Miwa, M. Jimbo, Holonomic Quantum Fields III, {\it RIMS, Kyoto Univ.}, 1979, V.15, 577--629

\bibitem{ZahMan}  V. E. Zakharov, S.V. Manakov, Construction of multidimensional integrable nonlinear systems and their solutions, 
 {\it Funk. Anal. Pril.}, 1985, V.19, N 2, 11--25
 
\bibitem{Dbar}   B.G. Konopelchenko, Solitons in Multidimensions. Inverse spectral transform  method, World Scientific, Singapore, 1993;  L.V. Bogdanov, V.E. Zakharov, On some developments of the di-bar dressing method, {\it Algebra and Analysis}, 1994, V.6, N.3, 40--58;  L.V. Bogdanov, Lingling Xue, A  class of reductions of the two-component KP hierarchy and the Hirota-Ohta system, {\it Theoret. and Math. Phys.},  2022, V.211, N.1, 473--482.


\end{thebibliography}
\end{document}